\documentclass{PoS}

\usepackage{transparent}
\usepackage{amsmath}
\usepackage{amssymb}
\usepackage{bm}
\usepackage[table,x11names]{xcolor}
\usepackage{graphicx}
\usepackage{gensymb}
\usepackage[font=footnotesize,labelfont=bf]{caption}
\usepackage{subcaption}

\title{Searching for TeV Gamma-ray Emission from Binary Systems with HAWC}

\ShortTitle{TeV Gamma-ray Binary Systems}

\author{\speaker{Chang Dong Rho}\\
        University of Rochester\\
        E-mail: \email{crho2@ur.rochester.edu}}
\author{Ryan Rubenzahl \& Segev BenZvi\\
		University of Rochester}

\author{For the HAWC Collaboration\\
		For a complete author list, see www.hawc-observatory.org/collaboration/icrc2017.php.}

\abstract{Compact binary systems can provide us with unique information on astrophysical particle acceleration and cosmic ray production. However, only five binary systems have ever been observed in TeV $\gamma$ rays. The High Altitude Water Cherenkov (HAWC) Observatory has high uptime (duty cycle $>95\%$) and a wide field of view (2 sr), making it well-suited for observing transient sources such as binaries. Using two years of data from HAWC, we have searched for TeV emission from three known TeV binary systems in the field of view and twenty-eight TeV binary candidates. We have searched the HAWC data for evidence of orbital modulation or flares from these objects, and report estimates of their $\gamma$-ray flux.}

\FullConference{35th International Cosmic Ray Conference --- ICRC2017\\
		10--20 July, 2017\\
		Bexco, Busan, Korea}

\begin{document}

\section{Introduction}\label{sec:intro}
Binaries that emit $\gamma$ rays are compact Galactic objects in orbit with massive companion stars. There are hundreds of X-ray binaries but only a few known $\gamma$-ray binaries \cite{dub13,seg15} and they are not well-understood. The production of $\gamma$ rays is suspected to arise when relativistic particles are accelerated at the shocks between the winds of a compact object and a massive star, or at the disk and the relativistic jets of a microquasar. Having such a low population of known TeV binaries, and the unexplained mismatches between emissions across different energy bands \cite{dub13} motivate a long-term search for new TeV binary systems. TeV binary systems have been observed to produce both periodic and flaring emission \cite{acc11,aha05,ale12} and they provide a unique resource for multiwavelength studies. However, their transient emission makes observations with pointed instruments a challenge.

Our analysis studies 28 $\gamma$-ray binary candidates selected from short period X-ray binaries (both low mass and high mass X-ray binaries) and 3 known $\gamma$-ray binaries in the Northern Hemisphere. We report either the maximum likelihood differential flux or the 95\% upper limits for each of the sources depending on their post-trial significances, using time integrated 25 months of HAWC data. Further analyses on a TeV $\gamma$-ray binary source HESS J0632+057 is also presented.

\section{HAWC}\label{sec:hawc}
The High Altitude Water Cherenkov (HAWC) Observatory is a ground array located at latitude of 19\degree N and at an altitude of 4,100 meters in Sierra Negra, Mexico. HAWC consists of 300 water Cherenkov detectors (WCDs) covering a large effective area of 22,000 m$^2$. Of the 300 deployed tanks, 294 have been instrumented \cite{hawccrabpaper}. Each WCD has a light-tight polypropylene bladder filled with 200,000 litres of purified water. The bladder is encased in a steel tank. At the bottom of each WCD there are three 8-inch Hamamatsu R5912 photomultiplier tubes (PMTs) oriented in an equilateral triangle and one 10-inch R7081-HQE PMT anchored at the center.

By combining the location and the time of each PMT triggered by an air shower, the core position and the angle at which the primary particle has generated the air shower is reconstructed to locate and identify the primary particle type. Simple topological cuts are applied to discriminate the air showers produced by hadronic cosmic rays from $\gamma$-ray air showers. For a detailed explanation of the event reconstruction, see \cite{smi15}.

The air shower trigger rate of HAWC is approximately 25kHz, more than 99.9\% originating from cosmic rays. At HAWC's altitude, a vertical shower from a 1 TeV photon will have about 7\% of the original photon energy when the shower reaches the tanks. This ratio increases to around 28\% at 100 TeV \cite{hawccrabpaper}. The main source of background to $\gamma$-ray observation is the hadronic cosmic-rays. Therefore, individual $\gamma$-ray-induced air showers are distinguished from cosmic-ray showers using their topology. HAWC has a duty cycle $>95\%$ and a wide, unbiased field of view of ~2 sr. As such, HAWC is well-suited to study long-duration light curves of astronomical objects, making it an excellent detector to search for $\gamma$-ray binary sources \cite{seg15}.

\section{Results}\label{sec:res}

\subsection{Update on TeV Binary Candidates}\label{ssec:update31}
Table 1 shows a list of point source fit results of the 31 TeV binary candidates. If the maximum likelihood test statistic \cite{you15} (TS) is $>10$, corresponding to a $\gtrsim2\sigma$ excess after trials, then the maximum likelihood flux at 7~TeV is reported. Otherwise, a 95\% upper limit is reported for the flux. A power law spectrum is assumed for point source fits and spectral index is fixed at -2.7. This value comes from the average index of all the resolved point sources in the 2-year HAWC catalog \cite{abe17} and the pivot energy of 7~TeV provides the least correlation between flux normalization and spectral index.

\begin{table}[ht]
\centering
\footnotesize
\begin{tabular}{l|r|r|l|r|r|r|r|r}
\multicolumn{1}{c|}{\bf Source} &
\multicolumn{1}{c|}{\bf RA} &
\multicolumn{1}{c|}{\bf Dec} &
\multicolumn{1}{c|}{\bf Type} &
\multicolumn{1}{c|}{$\mathbf{d}$} &
\multicolumn{1}{c|}{$\bm{\tau}$} &
\multicolumn{1}{c|}{$\mathbf{TS}$} &
\multicolumn{1}{c|}{\bf Signif.} &
{\bf Dif Flux @ 7~TeV} \\
 & & & & \multicolumn{1}{c|}{[kpc]} & 
 \multicolumn{1}{c|}{[day]} & &
 [post-trial] & 
 [TeV$^{-1}$ cm$^{-2}$ s$^{-1}$] \\
\hline
IGR J00370+6122        &  00:37 & $+61^\circ21'$ & HMXB  &  3.4 & 15.67     &  0.00 & $<2\sigma$ &  $9.21\cdot10^{-15}$ \\
V662 Cas               &  01:18 & $+65^\circ17'$ & HMXB  &  6.5 & 11.60     &  0.15 & $<2\sigma$ &  $5.75\cdot10^{-14}$ \\
IGR J01363+6610        &  01:36 & $+66^\circ11'$ & HMXB  &  2.0 & --        &  0.00 & $<2\sigma$ &  $6.03\cdot10^{-14}$ \\
IGR J01583+6713        &  01:58 & $+67^\circ13'$ & XB    &  4.1 & --        &  0.26 & $<2\sigma$ &  $9.51\cdot10^{-14}$ \\
VES 737                &  02:20 & $+63^\circ01'$ & Bin   &  5.0 & --        &  0.26 & $<2\sigma$ &  $4.38\cdot10^{-14}$ \\
\rowcolor{LavenderBlush2}
LS I +61 303           &  02:40 & $+61^\circ13'$ & HMXB  &  2.0 & 26.50     &  0.00 & $<2\sigma$ &  $1.66\cdot10^{-14}$ \\
XTE J0421+560          &  04:19 & $+55^\circ59'$ & HMXB  &  2.0 & 19.41     &  0.04 & $<2\sigma$ &  $1.29\cdot10^{-14}$ \\
GRO J0422+32           &  04:21 & $+32^\circ54'$ & LMXB  &  2.0 &  0.21     &  1.43 & $<2\sigma$ &  $3.93\cdot10^{-15}$ \\
RX J0440.9+4431        &  04:40 & $+44^\circ31'$ & HMXB  &  2.9 & --        &  7.41 & $<2\sigma$ &  $1.10\cdot10^{-14}$ \\
IGR J06074+2205        &  06:07 & $+22^\circ05'$ & HMXB  &  4.5 & --        &  0.01 & $<2\sigma$ &  $2.15\cdot10^{-15}$ \\
V616 Mon               &  06:22 & $-00^\circ20'$ & LMXB  &  1.1 &  0.33     &  0.00 & $<2\sigma$ &  $2.54\cdot10^{-15}$ \\
\rowcolor{LavenderBlush2}
HESS J0632+057         &  06:32 & $+05^\circ48'$ & HMXB  &  1.6 & $315\pm5$ &  2.39 & $<2\sigma$ &  $4.85\cdot10^{-15}$ \\
PSR J1023+0038         &  10:23 & $+00^\circ53'$ & LMXB   &  1.3 & --        &  5.27 & $<2\sigma$ &  $6.83\cdot10^{-15}$ \\
XTE J1118+480          &  11:18 & $+48^\circ02'$ & LMXB    &  1.7 &  0.17     &  1.84 & $<2\sigma$ &  $9.71\cdot10^{-15}$ \\
LS IV -01 1            &  17:07 & $-01^\circ05'$ & Star  &  0.3 & --        &  0.00 & $<2\sigma$ &  $1.68\cdot10^{-15}$ \\
PSR J1810+1744         &  18:10 & $+17^\circ41'$ & MSP   &  2.0 & --        &  0.25 & $<2\sigma$ &  $2.62\cdot10^{-15}$ \\
PSR J1816+4510         &  18:16 & $+45^\circ10'$ & MSP   &  4.0 &  0.36     &  0.15 & $<2\sigma$ &  $5.53\cdot10^{-15}$ \\
\rowcolor{Thistle3}
LS 5039                &  18:26 & $-14^\circ50'$ & HMXB  &  2.9 &  3.90     & 139.97 & $11.54\sigma$ &  $6.37\cdot10^{-14}$ \\
\rowcolor{Azure2}
4U 1907+09             &  19:09 & $+09^\circ49'$ & HMXB  &  4.0 &  8.37     & 10.87 & $2.17\sigma$ &  $7.08\cdot10^{-15}$ \\
\rowcolor{Azure2}
SS 433                 &  19:12 & $+04^\circ59'$ & XB    &  5.5 & 13.10     & 17.27 & $3.29\sigma$ &  $8.51\cdot10^{-15}$ \\
\rowcolor{Azure2}
IGR J1914+0951         &  19:14 & $+09^\circ52'$ & HMXB  &  5.0 & 13.56     & 80.40 & $8.58\sigma$ &  $1.50\cdot10^{-14}$ \\
Cyg X-1                &  19:58 & $+35^\circ12'$ & HMXB    &  2.2 &  5.60     &  4.99 & $<2\sigma$ &  $5.96\cdot10^{-15}$ \\
PSR J1959+2048         &  19:59 & $+20^\circ48'$ & Bin   &  2.5 & --        &  3.10 & $<2\sigma$ &  $4.08\cdot10^{-15}$ \\
GS 2000+251            &  20:02 & $+25^\circ14'$ & LMXB    &  2.7 &  0.35     &  0.00 & $<2\sigma$ &  $2.14\cdot10^{-15}$ \\
V404 Cyg               &  20:24 & $+33^\circ52'$ & LMXB    &  2.4 &  6.47     &  0.98 & $<2\sigma$ &  $3.97\cdot10^{-15}$ \\
\rowcolor{Azure2}
EXO 2030+375           &  20:32 & $+37^\circ38'$ & HMXB  &  5.0 & 46.02     & 16.85 & $3.23\sigma$ &  $9.54\cdot10^{-15}$ \\
\rowcolor{Azure2}
Cyg X-3                &  20:32 & $+40^\circ57'$ & HMXB    &  7.0 &  0.20     & 77.54 & $8.41\sigma$ &  $2.15\cdot10^{-14}$ \\
LS III +49 13          &  20:56 & $+49^\circ40'$ & BH    &  0.1 & --        &  0.00 & $<2\sigma$ &  $4.67\cdot10^{-15}$ \\
SAX J2103.5+4545       &  21:03 & $+45^\circ45'$ & HMXB  &  6.5 & 12.68     &  0.00 & $<2\sigma$ &  $2.43\cdot10^{-15}$ \\
4U 2206+543            &  22:07 & $+54^\circ31'$ & HMXB  &  2.6 &  9.57     &  0.00 & $<2\sigma$ &  $3.36\cdot10^{-15}$ \\
MWC 656                &  22:42 & $+44^\circ43'$ & HMXB  &  2.6 & --        &  0.40 & $<2\sigma$ &  $6.10\cdot10^{-15}$ \\
\end{tabular}
\caption{Observations of TeV binary candidates. Known TeV binaries \cite{dub13} are highlighted in red. Candidates with $TS>10$ (significance~$>2\sigma$ after trials) are highlighted in blue. Note that if the post-trials significance is $<2\sigma$ the reported flux corresponds to a 95\% UL.}
\label{table:results}
\end{table}

\begin{figure}[!htb]
  \centering
  \includegraphics[width=0.8\textwidth]{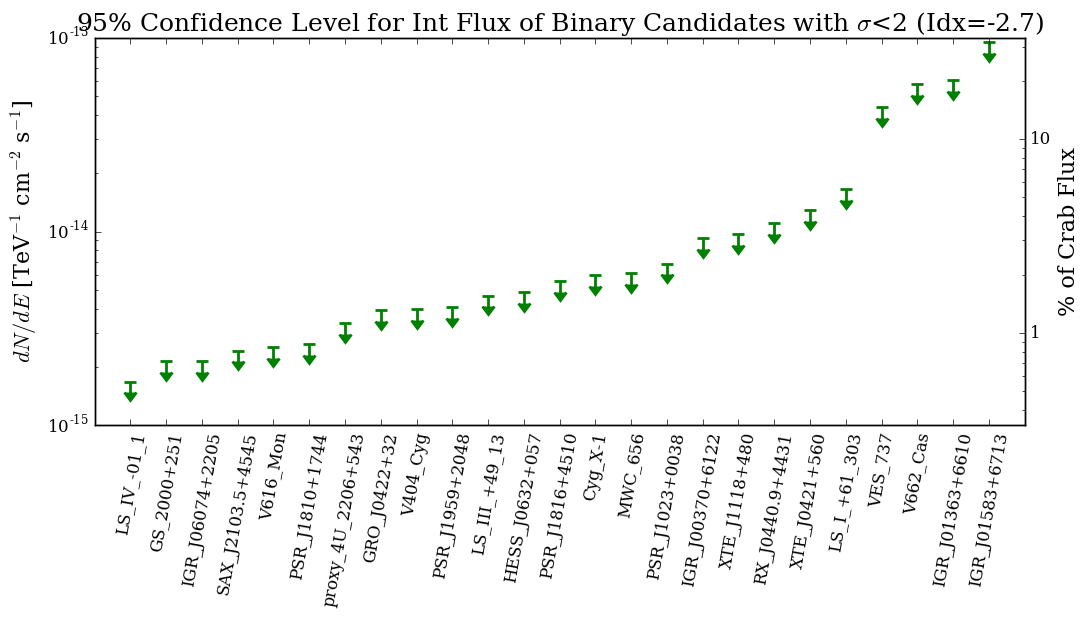}
  \put(-300,160){\huge\transparent{0.2}\color{red}{Preliminary}}
  \caption{\sl 95\% differential flux upper limits at 7 TeV from candidates with $<2\sigma$ significance after trials. The upper limit is a Feldman-Cousins interval \cite{felcou}. \label{fig:binUL}}
\end{figure}

\begin{figure}[!htb]
  \centering
  \includegraphics[width=0.8\textwidth]{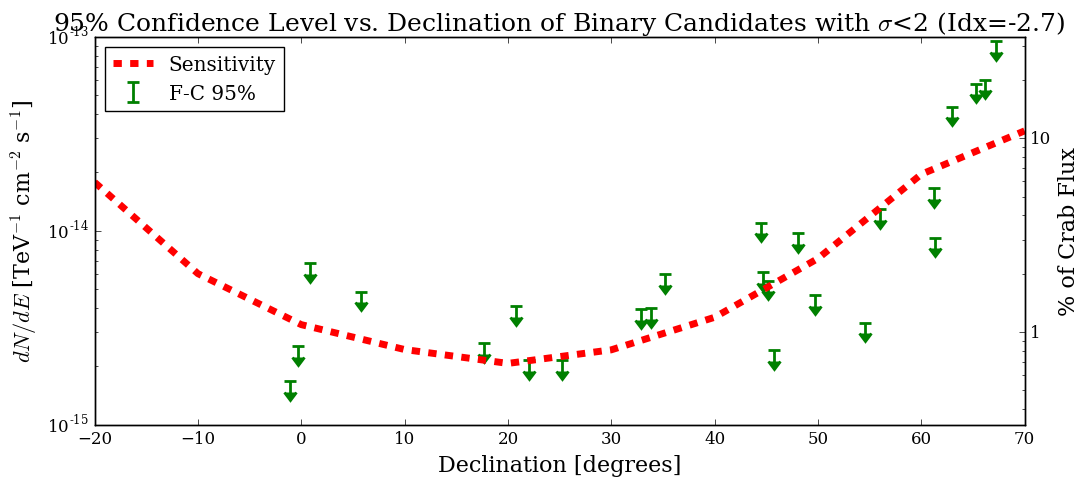}
  \put(-220,100){\huge\transparent{0.2}\color{red}{Preliminary}}
  \caption{\sl 95\% flux upper limits at 7 TeV from Figure~\ref{fig:binUL} plotted as a function of declination. The red dotted line shows HAWC's sensitivity curve at the index of 2.7, where sensitivity is defined as the median upper limit without signal. \label{fig:binULdec}}
\end{figure}

Most of the binary candidates do not have statistically significant excess counts. The flux upper limits for sources without significant excesses are plotted in Figure~\ref{fig:binUL}. The upper limits were compared against their positions in declination as shown in Figure~\ref{fig:binULdec}. The sources with higher upper limits are influenced by the sensitivity of the detector, which varies with declination.

Of the 31 objects investigated, six candidates have excesses $>2\sigma$ after trials, and three of them have excess counts with $>5\sigma$ significance. Note that all of these objects are in between the 18h and 20h right ascension bands surrounded by a number of other pointlike and extended sources along the Galactic Plane. Also, they are likely to be affected by Galactic diffuse emission.

For the six candidates with excesses $>2\sigma$, light curves were plotted but no orbital modulations of flux were observed using daily maps containing 17 months of HAWC data. The daily maps spanning up to 25 months are under development. While we are attempting to carry out multiple source fits in our regions of interest, including Galactic diffuse emission, we are unable to definitively associate the excess counts with the binary candidates at this time. All the significant excesses remain consistent with contamination from nearby extended sources.

In the following section, a detailed analysis on one of the sources, HESS J0632+057, is presented.

\subsection{HESS J0632+057}\label{ssec:res0632}
HESS J0632+057 was first detected as a TeV source by the HESS collaboration in 2007 \cite{aha07}. The nearby Be star, MWC 148, suggested the possibility that HESS J0632+057 is a binary source. An orbital modulation of its flux was later observed in the X-ray band by Swift-XRT with an orbital period at 321$\pm$5 days \cite{bon11}. This identification of HESS J0632+057 as a binary system was the first time that a system which had initially been identified as a TeV $\gamma$-ray observation was later recognized as a binary source. Orbital modulation of the TeV emission has also been confirmed in the $\gamma$-ray band by VERITAS in 2016, which measured a period of 315$\pm$5days \cite{sch16}. HAWC data have been used to look for a similar orbital modulation of flux with daily maps ranging 17 months. This, however, was not successful. Hence, 95\% upper limits for energy spectrum were calculated using HAWC data and a combined data plot of VERITAS and HAWC is presented for HESS J0632+057.

\begin{figure}[!htb]
  \centering
  \begin{subfigure}[b]{0.45\textwidth}
  \includegraphics[width=\linewidth]{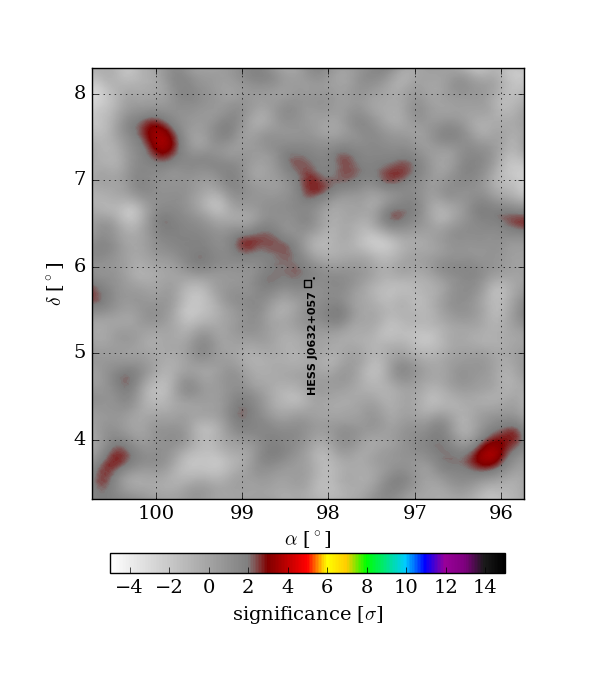}
  \put(-140,180){\huge\transparent{0.2}\color{red}{Preliminary}}
  \end{subfigure}
  \caption{Significance map of HESS J0632+057 for the 25 month dataset. TeVCAT labels have been added to indicate the known location of HESS J0632+057.}
  \label{fig:sig0632}
\end{figure}

Figure~\ref{fig:sig0632} is the significance map at the reported location of HESS J0632+057. With labels from an online catalog for TeV Astronomy (TeVCAT, Scott Wakely \& Deirdre Horan), it is evident to be isolated in the region and the whole area does not display much excess. The pre-trial significance at (RA : 98.24\degree, Dec : 5.81\degree) in this plot is 1.63$\sigma$.

\begin{figure}[!htb]
  \centering
  \includegraphics[width=1.0\textwidth]{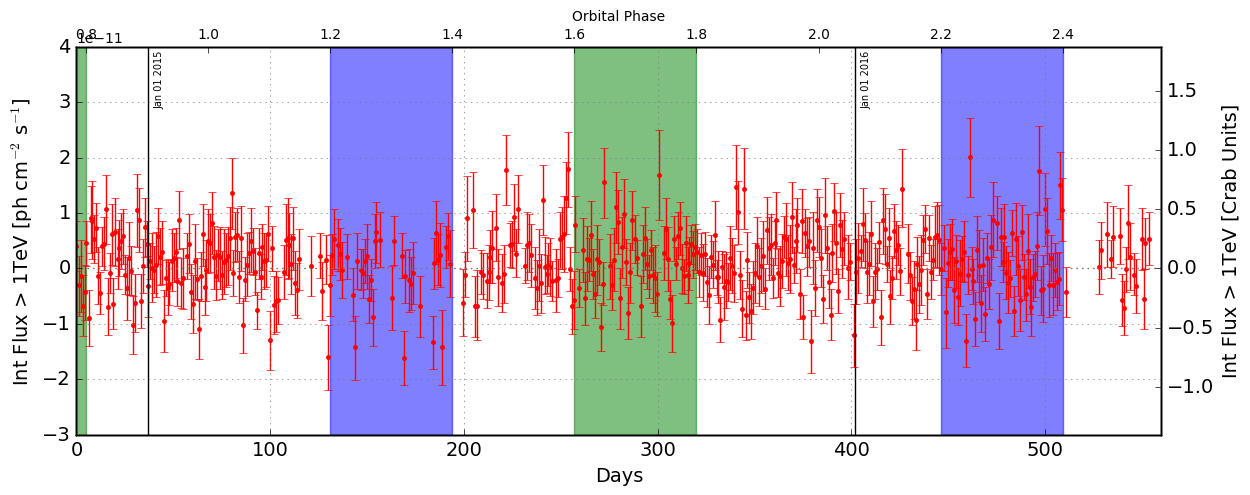}
  \put(-370,135){\huge\transparent{0.2}\color{red}{Preliminary}}
  \caption{Light curve of HESS J0632+057 using 17 month HAWC dataset. Highlighted regions indicate orbital phases between: 0.2 - 0.4 (blue); 0.6 - 0.8 (green).}
  \label{fig:lc0632}
\end{figure}

Figure~\ref{fig:lc0632} is the light curve produced with 17 months of HAWC data for HESS J0632+057 \cite{lightcurve}. The two highlighted regions indicate high state orbital phases of 0.2 to 0.4 (blue) and 0.6 to 0.8 (green), obtained from observations with VERITAS \cite{mai15}.

\begin{figure}[!htb]
  \centering
  \includegraphics[width=1.0\textwidth]{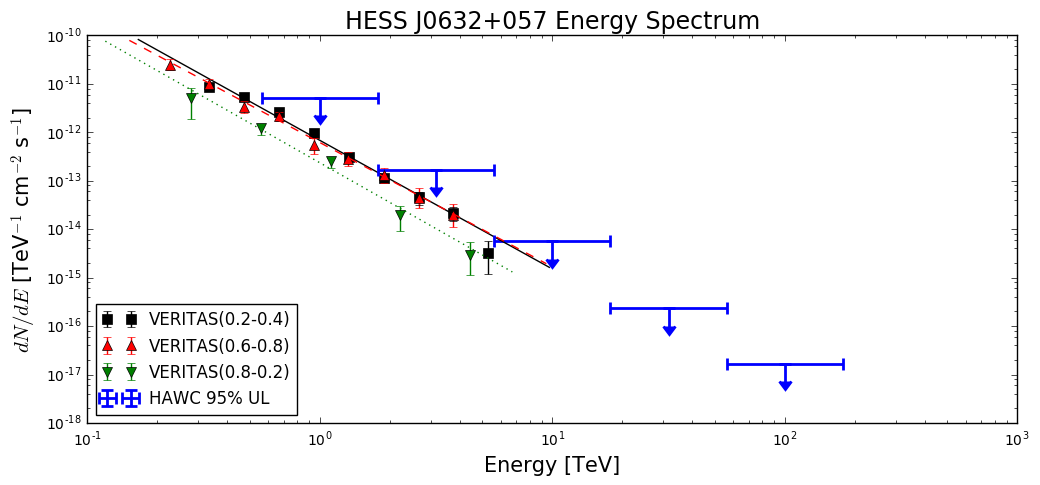}
  \put(-115,165){\huge\transparent{0.2}\color{red}{Preliminary}}
  \caption{\sl Combined results of VERITAS \cite{sch16} and HAWC for HESS J0632+057. Black squares are the VERITAS observation in the orbital phases between 0.2 and 0.4. Red Triangles are between 0.6 and 0.8 (the two high states). In green is the low state phase between 0.8 and 0.2. Finally, blue is HAWC's 95\% upper limits at various energies calculated using Feldman-Cousins confidence level with a power law spectrum of index 2.7.
  \label{fig:verhaw0632}}
\end{figure}

Figure~\ref{fig:verhaw0632} shows the spectrum measured by VERITAS and upper limits calculated with data from HAWC. The upper limits range from 1 TeV to 100 TeV.

\section{Discussion}\label{sec:disc}
We have investigated the 28 TeV binary candidates and the 3 known binary systems with 25~months of data from HAWC and have presented 95\% upper limits for those with post-trial significances $<2\sigma$. While a few candidates (LS 5039, 4U 1907+09, SS 433, IGR J1914+0951, EXO 2030+375 and Cyg X-3; Table 1) exhibit significant excesses after trials, these excesses cannot be disentangled from fluctuations in the background, confused emission from nearby sources, or diffuse emission. We have not yet observed any periodic signatures or flares in their fluxes. Also, the calculated upper limits (Figure~\ref{fig:binUL}) were reordered by their positions in declination to look for a possible association of higher upper limits with higher declinations where HAWC's sensitivity is relatively poor (Figure~\ref{fig:binULdec}). The general distribution of the upper limits follow the expected sensitivity curve in red and the 4 candidates with the highest upper limits were all positioned above $60\degree$ in declination.

HESS J0632+057 has been studied in more depth. Searches for high states using the light curve were attempted (Figure~\ref{fig:lc0632}) but were not successful as no clear high states were observed in the highlighted regions. 95\% upper limits from HAWC were calculated (Figure~\ref{fig:verhaw0632}). The derived upper limits are slightly higher than the observations from VERITAS. There is also a possibility that the energy spectrum of HESS J0632+057 cuts off above a few TeV, which may be hinted by the highest energy VERITAS point in Figure~\ref{fig:verhaw0632} during the orbital phase between 0.2 and 0.4 (black).

\section{Conclusion}\label{sec:conc}
Upper limits are presented for HESS J0632+057 using 25 months of data from HAWC. The non-observation does not contradict the VERITAS measurements since the HAWC upper limits are above the VERITAS data points. By extrapolating the VERITAS data in Figure~\ref{fig:verhaw0632}, HAWC is expected to have the significance of $5\sigma$ in about 3 years to observe the high states in the orbital phase between 0.2 and 0.4. If the energy spectrum cuts off above 5 TeV, it would take about 8 years to reach this significance. 

For the 31 TeV binary candidates, their post-trial significances were calculated and the upper limits were calculated for the less significant 26 sources. Upon comparing with HAWC's sensitivity, the distribution of 95\% confidence intervals was similar to the sensitivity.

HAWC has not observed a TeV binary source \cite{abe17}. Part of the reason may be source confusion. The likelihood analysis is capable of performing multiple source fits to reduce source confusion. However, due to the significant degrees of freedom involved and the requirement for a good understanding of a Galactic diffuse model and a proper identification of nearby sources, this method was not used for the 31 TeV binary candidate analysis. However, we expect multi-source fitting (including Galactic diffuse emission) will be applied in the future analysis.

{}

\end{document}